\documentclass{PoS}
\usepackage{graphicx}

\title{PRODUCTION OF FAST NEUTRON WITH A PLASMA FOCUS DEVICE}

\ShortTitle{PLASMA FOCUS NEUTRON}

\author{Moshe Gai \thanks{Work Supported by USDOE Grant No. DE-FG02-94ER40870.}\\

       Laboratory for Nuclear Science at Avery Point, 
University of Connecticut, \\ 1084 Shennecossett Rd, Groton, CT 06340-6097.\\
\hspace{1cm} and\\
Department of Physics, WNSL Rm 102, Yale University, PO Box 208124, \\
272 Whitney Avenue, 
New Haven, CT 06520-8124.\\
\       \\
http://astro.uconn.edu, E-mail: \email{moshe.gai@yale.edu}}

\abstract{Before its demise DIANA Hi-TECH, LLC, demonstrated the use of two 50 kJoule Plasma Focus devices for the copius production of fast neutrons, x-rays and radio-isotopes. Such a device is suitable for fast neutron non invasive interogation of contra-band materials including hidden nuclear materials. It could be particularly useful for a {\bf fast} and fail safe interogation of large cargo containers, or in merchant marine port of entries. The performance and fast neutron production (2.5 or 14 MeV at $10^{11}$ or $10^{13}$ per pulse, respectively) of the two PF50 Plasma Focus devices produced by DIANA HiTECH, LLC, are discussed.}

\FullConference{International Workshop on Fast Neutron Detectors and Applications \\

		 April, 3 - 6, 2006\\

		 University of Cape Town, South Africa}

\begin{document}

\section{Introduction}
         The Plasma Focus (PF) phenomenon was independently
discovered by J. Mather \cite{Mat65} and N. V.Filippov \cite{Fil62} in the
late 50's. Since then advances were made \cite{IAEA} in many laboratories that studied the
capabilities of PF for producing short pulses (10 - 100 ns) of
X-rays, neutrons and fast ions depending on PF mode of
operation. The production of radio-isotopes in PF has recently received 
         attention \cite{DIANA,ANG05}. Commercial PF devices are now available from All-Russia
         Research Institute of Automatics \cite{ARRI}. PF operation can be briefly described as a
five stage process (see Fig. 1) with the appearance of
high-energy ions and nuclear reactions occurring in
the last stage. In Fig. 2 we show a picture of the PF25 operated at DIANA HiTECH, LLC, in 
         North Bergen, New Jersey,.
 
These studies \cite{Nar88} suggest the following empirical formula 
for the flux of ions with (E $>$ 0.1 MeV):
$d\Phi / dE \ \approx \ E^{-m} \ (2<m<3)$. 
This spectral dependence of ions trapped in the plasma 
leads to the observed neutron spectrum emitted from a 
plasma focus device operating with deuterium gas. The data provide evidence
that the yield of fast beams as well as the reaction yields
in plasma ($Y_p$) and/or with external targets ($Y_t$) placed inside a PF, all scale with
the square cf the energy (W) stored in the capacitor
bank \cite{Brz97}. 

 \begin{figure}
\hspace{1cm} \includegraphics[width=4in]{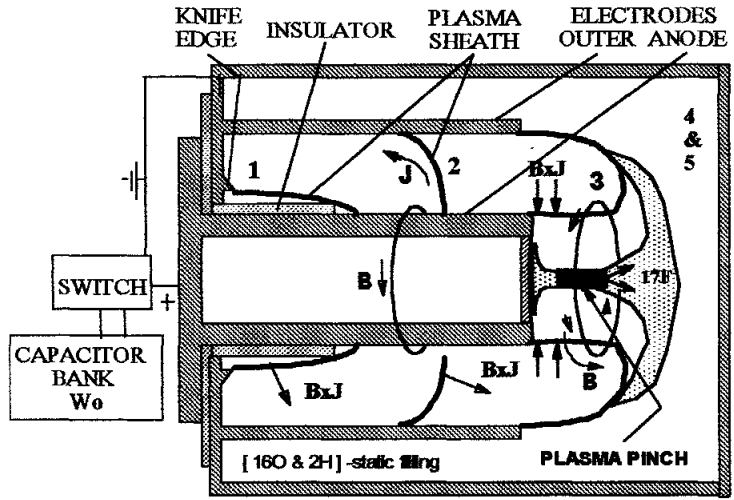}
\vspace{-5mm} \caption{\label{PF} Schematic operation of a plasma focus
header (used for the production of $^{17}F$).
 We indicate the sequence of plasma sheath positions. The
central electrode diameter is approximately 50 mm, the chamber is filled
with oxygen plus deuterium gas mixture at p = 0.1 - 7 Torr. The plasma
develops in stages: 1 - the plasma sheath is formed, 2 - the
plasma sheath moves toward the anode nozzle (v $\approx \ 10^5$ m/s), 
         3 - the sheath arrives at the end of the anode and rearranges 
into a cylinder with a conical opening, 4 - the plasma is
compressed at the axis ($10^{23} \  ions/m^3$), 5 - the plasma column
quickly develops instabilities associated with high energy acceleration 
        (with large nuclear reactivity and intense X-ray emission).
 Small (20 -  300 microns) plasma domains are created. The
domains have above solid state densities, and temperature of kT $>$ 3 keV and
magnetic fields sufficient to trap ions with kinetic energy of up to 5 MeV/nucleon.}
\end{figure} 
 
 \begin{figure}
\hspace{1.5In} \includegraphics[width=3in]{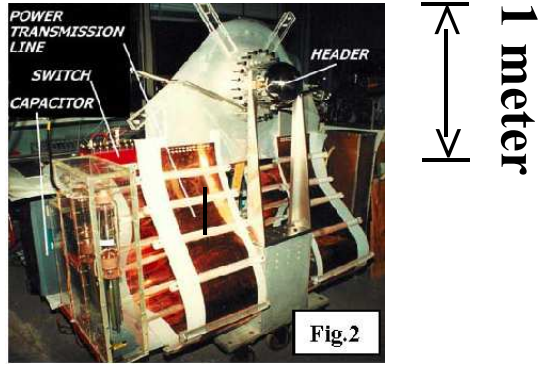}
 \vspace{-6mm}
 \caption{\label{PF2} A picture of the PF25 operated at DIANA, HiTECH, LLC.}
\end{figure} 

 \begin{figure}
\hspace{1In} \includegraphics[width=3.5in]{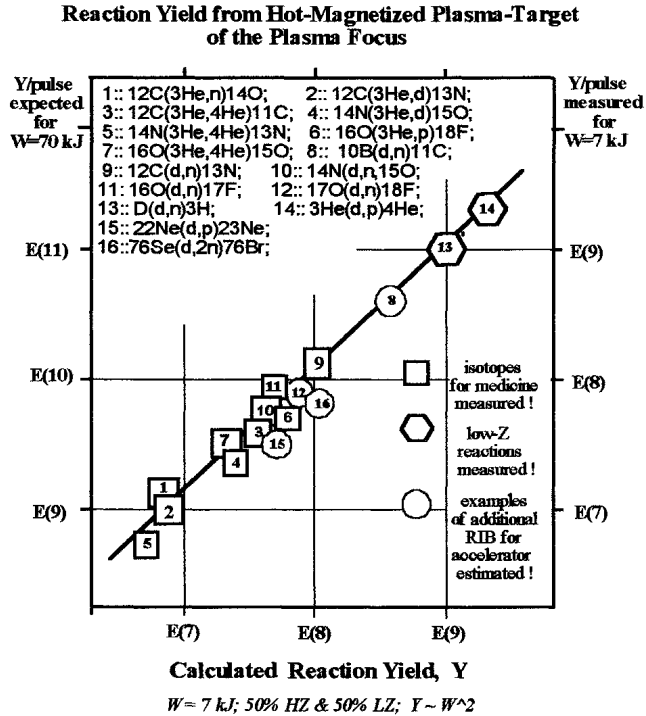}
 \vspace{-6mm}
 \caption{\label{PF3} Measured production rates (with PF7) and predicted 
 yield (for PF70). Neutrons (2.5 MeV) are produced in PF loaded with deuterium gas (13). 
 For a mixture of d + t the production rate of 14 Mev neutrons is 100 times larger.}
\end{figure} 

\section{The Steven Tech and DIANA's PF5, PF25 and PF50}

The experimental group at Steven Tech has achieved high
efficiency (yield per kJ of stored energy) for the
production of short-lived radioisotopes in a plasma
environment \cite{Brz91} using a Mather-type PF device with a
capacitor bank energy of W= 7 kJ (18,000 Volts) and a 
gas mixture pressure of p $\approx$ 5 Torr. Gas mixtures
were composed of low-Z (LZ) isotopes (hydrogen or deuterium) 
mixed with high-Z (HZ) isotopes.
The obtained results are shown in Fig. 3, from which it is 
clear that a broad range of
radioisotopes, with large yield, can be produced in a PF
device. A relatively small PF-machine, operated at a
discharge power of W= 7 kJ, produced $10^6$ to 
$5\times10^8$ radio-nuclei per pulse, while with a medium size
machine, operated at discharge power of, W = 70 kJ,
one can expect a hundred times higher yields.

\section{Production of Fast Neutrons}

Of particular interest are PF devices operating with a deuterium gas or 
deuterium plus tritium gas mixture. They yield 2.5 and 14 MeV fast neutrons, 
respectively. Since the cross section for t(d,n)$\alpha$ reaction is 100 times larger than 
d(d,n) reaction, the production rate of 14 MeV neutrons is 100 times larger under 
the same PF conditions. The rates of neutrons produced in two PF50 devices
operating with deuterium gas were measured 
at DIANA HiTECH, LLC, by measuring the activation of silver by the moderated neutrons 
to produce $^{108}Ag$ ($\tau$ =  2.41 min) and $^{110}Ag$ ($\tau$ = 24.6 sec). The activation 
method was calibrated using an intense Pu-Be source at the Wright Laboratory at Yale University.
The obtained neutron production rate with the two PF50 device tested at DIANA HiTECH, LLC,
 confirmed (after scaling) the production rate predicted for the 
 PF70 device ($10^{11}$ n/pulse for d-d fusion and $10^{13} $
n/pulse for d-t fusion) as shown in Fig. 3. The two 
PF50 devices were operated over a long period (a few days) at a rate of 1 Hz. The control 
system of the PF was shielded and contained in a Faraday cage due to the 
intense associated electromagnetic pulse.

\end{document}